\documentclass{ws-procs9x6}

\begin{document}

\title{Interactions of ultrahigh-energy neutrinos} 

\author{Alexander Kusenko}

\address{Department of Physics and Astronomy, UCLA, Los Angeles, CA
90095-1547\\ RIKEN BNL Research Center, Brookhaven National
Laboratory, Upton, NY 11973
}

\maketitle

\abstracts{ Future detection of  ultrahigh-energy neutrinos will
  open a new window on physics at center-of-mass energy $10^5$~GeV and 
  higher. In particular, observations of neutrino-initiated showers 
  will help test the Standard Model predictions for the
  neutrino-nucleon cross section.  
}

The anticipated detection of ultrahigh-energy (UHE) neutrinos will mark the
beginning of UHE neutrino astronomy and will provide new opportunities for
particle physics.  The prospects for detection of 
upgoing neutrinos by ICE CUBE~\cite{ice}, 
as well as the ground-level fluorescence
detectors~\cite{Feng}, such as HiRes and Pierre Auger~\cite{Bertou},
and orbiting detectors~\cite{Domokos,Fargion}, such as EUSO,  
present an
opportunity to conduct a {\em particle physics experiment}~\cite{kw} and to 
measure the neutrino-nucleon cross section $\sigma_{\nu_N}$ at an
unprecedented center-of-mass energy $10^5-10^6$~GeV.  The relative rates of
the horizontal air showers (HAS) and UAS initiated by neutrinos depend on
$\sigma_{\nu_N}$ in such a way that the cross section can be determined
without a precise knowledge of the incident neutrino flux.  Moreover, the
angular distribution of UAS provides an additional and independent
information about the cross section.

There are several reliable predictions regarding the flux of UHE neutrinos.
Observations of ultrahigh-energy cosmic rays (UHECR) imply the existence of
a related flux of ultrahigh-energy neutrinos generated in the interactions
of UHECR with cosmic microwave background radiation~\cite{UHECR}.  In
addition, active galactic nuclei~\cite{AGN}, gamma-ray bursts~\cite{GRB},
and other astrophysical objects can produce a large flux of
neutrinos~\cite{Waxman}.  Finally, some of the proposed explanations of the
puzzle~\cite{UHECR} of UHECR predict a strong additional flux of
ultrahigh-energy neutrinos~\cite{Zburst1,Zburst2,kalashev}.  The flux of
UHE neutrinos at energies $10^{18}-10^{20}$~eV is uncertain.  However, as
discussed below, the proposed measurement of the neutrino cross section is
not very sensitive to these uncertainties~\cite{kw}.

Calculations of the neutrino-nucleon cross section $\sigma_{\nu_N}$ at
$10^{20}$~eV necessarily use an extrapolation of parton distribution
functions and Standard Model parameters far beyond the reach of present
experimental data.  The resulting cross section~\cite{UHEnusig} at
$10^{20}$~eV is $\sim 10^{-31}{\rm cm}^2$.  It is of great interest to
compare this prediction with experiment to test the small-$x$ behavior of
QCD, as well as the possible contributions of new physics beyond the
electroweak scale~\cite{basu}.

For the purposes of such a measurement, we assume the cross
section to be a free parameter bounded from below by the value $\sim
2\times 10^{-34}{\rm cm}^2$ measured at HERA at $\sqrt{s}=314$~GeV. (This 
corresponds to a laboratory energy $E_\nu=5.2\times 10^{13}$~eV of an
incident neutrino.)

UHE neutrinos are expected to arise from pion and muon decays.  The subsequent
oscillations generate a roughly equal fraction of each neutrino flavor. Tau
neutrinos interacting below the surface of the Earth can create an
energetic $\tau$-lepton, whose decay in the atmosphere produces an UAS.

It is clear that, for smaller values of the cross section, the Earth is
more transparent for neutrinos, so that more of them can interact just
below the surface and produce a $\tau$ that can come out into the
atmosphere.  As long as the mean free path $\lambda_\nu $ is smaller than
the radius of the Earth, the rates of UAS increase with $\lambda_\nu
\propto 1/\sigma_{\nu_N} $.  The rates of HAS, however, are proportional to
$\sigma_{\nu_N}$; they decrease for a smaller cross section.  The
comparison of the two rates, shown in Fig.~1, can allow a measurement of
the cross section which is practically independent of the uncertainties in
the incident neutrino flux.

\begin{figure}[t]
\centering
\hspace*{-5.5mm}
\leavevmode\epsfysize=6cm 
\epsfbox{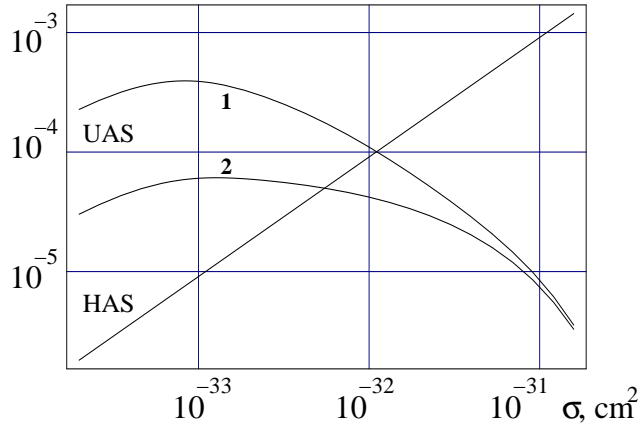} 
\caption[fig. 2]{\label{fig2} The air shower probability per incident tau
neutrino as a function of the neutrino
cross section. 
The incident neutrino energy is $10^{20}$~eV and the assumed 
energy threshold for detection of UAS is
$E_{\rm th}=10^{18}$eV for curve 1 and $ 10^{19}$eV for curve 2.
}
\end{figure} 

In addition, the angular distribution of UAS alone can be used as an
independent measurement of the cross section.  The peak of the angular
distribution of UAS occurs~\cite{kw} when $\cos \theta_{\rm peak} \approx 
\lambda_\nu/ 2 R_\oplus$, which depends on the cross section. 

It is comforting to know that the program of UHE neutrino astronomy, which
is one of the goals of EUSO and OWL, is not at risk, regardless of any
theoretical uncertainties in the neutrino cross section.  For a larger
cross section, HAS are more frequent than HAS, while for a smaller value
UAS dominate.  Nevertheless, the total rates of combined events remain
roughly constant for a wide range of $\sigma_{\nu_N}$, as shown in Fig.~1.

On the other hand, some of the reported bounds on the neutrino flux are
directly affected by the uncertainties in the neutrino-nucleon cross
section.  For example, the reported bounds on the UHE neutrino flux due to
the non-observation of neutrino-initiated HAS~\cite{FlyEye} and of radio
signals produced by neutrino interactions near the surface of the
moon~\cite{GLN99} are weaker if the cross section is smaller.

To summarize, the future neutrino experiments can determine the
neutrino-nucleon cross section at energies as high as $10^{11}$~GeV, or
higher, by comparing the rates of UAS with those of HAS; or by measuring
the angular distribution of UAS events.  Therefore, there is an exciting
opportunity do a particle physics experiment using neutrinos produced by
natural astrophysical sources.

\section*{Acknowledgments}

This work was supported in part by the DOE grant DE-FG03-91ER40662.

\end{document}